\documentclass{iaus}
\usepackage{graphicx}

\title[Constraining the Age and He Content of 47 Tuc]
{Globular Cluster Ages from Main Sequence Fitting and Detached, 
Eclipsing Binaries: The Case of 47 Tuc}

\author[A. Dotter, J. Kaluzny, \& I. B. Thompson]
{Aaron Dotter$^1$, Janusz Kaluzny$^2$, \and Ian B. Thompson$^3$}

\affiliation{$^1$Dept. of Physics \& Astronomy, University of Victoria\\ 
Victoria BC V8P 5C2 Canada \\ 
email: {\tt dotter@uvic.ca} \\[\affilskip]
$^2$Copernicus Astronomical Center\\ Bartycka 18,
00-716 Warsaw, Poland \\ email: {\tt jka@camk.edu.pl}\\
$^3$Carnegie Observatories\\ 813 Santa Barbara St.,
Pasadena, CA 91101-1292 \\email: {\tt ian@ociw.edu}}

\pubyear{2008}
\volume{258}  
\pagerange{111--222}
\jname{The Ages of Stars}
\editors{A.C. Editor, B.D. Editor \& C.E. Editor, eds.}

\begin{document}

\maketitle

\begin{abstract}
Age constraints are most often placed on globular clusters by 
comparing their CMDs with theoretical isochrones. The recent 
discoveries of detached, eclipsing binaries in such systems 
by the Cluster AgeS Experiment (CASE) provide new insights into 
their ages and, at the same time, much needed tests of
stellar evolution models. We describe efforts to model 
the properties of the detached, eclipsing binary V69 in 47 Tuc and 
compare age constraints derived from stellar evolution models of 
V69A and B with ages obtained from fitting isochrones to the 
cluster CMD. We determine, under reasonable assumptions 
of distance, reddening, and metallicity, the extent to which it is 
possible to simultaneously constrain the age and He content of 47 Tuc.
\end{abstract}


\section{Background and Previous Results}
47 Tuc is among the closest and most carefully studied globular
clusters in the Galaxy. Table \ref{tab1} lists the best available
estimates of [Fe/H], [$\alpha$/Fe], distance, and reddening along
with a representative sample of ages from recent studies.

\begin{table}
\begin{center}
\caption{47 Tuc: Basic Parameters and Previous Age Results}
\label{tab1}
\begin{tabular}{|c|c|c|}\hline 
{\bf Parameter} & {\bf Value} & {\bf Source} \\ \hline
[Fe/H] & --0.76$\pm$0.01$\pm$0.04 & Koch \& McWilliam (2008)\\
  & --0.75$\pm$0.01$\pm$0.04 & McWilliam \& Bernstein (2008)\\           
\hline
[$\alpha$/Fe]& $\sim$0.4 & Koch \& McWilliam (2008)  \\
             & $\sim$0.3 & McWilliam \& Bernstein (2008)\\
\hline
$DM_V$       & 13.35$\pm$0.08    & Thompson et al. (2009) \\ 
\hline
E(B--V)       & 0.0320$\pm$0.0004 & Schlegel et al. (1998)\\ 
\hline
Age (Gyr)& 10--13                  & Salaris et al. (2007) \\
         & $\sim$11.3              & Gratton et al. (2003) \\
         & 11.0$\pm$1.4            & Percival et al. (2002) \\
         & 10.7$\pm$1.0            & Salaris \& Weiss (2002)\\
         & 11.5$\pm$0.8            & VandenBerg (2000) \\
         & 12.5$\pm$1.5            & Liu \& Chaboyer (2000) \\
\hline
\end{tabular}
\end{center}
\end{table}

\section{Stellar Evolution Models}
The stellar evolution models utilized in this contribution
were computed using the Dartmouth Stellar Evolution Program (DSEP;
Dotter et al. 2007, 2008).  The models include the effects of 
partially inhibited microscopic diffusion of He and metals 
(Chaboyer et al. 2001).  The models were converted to the 
observational plane using PHOENIX synthetic spectra.  Stellar 
evolution tracks and isochrones were computed for [Fe/H]= --0.8, 
--0.75, and --0.7; [$\alpha$/Fe]= 0, +0.2, and +0.4; and Y=0.24, 
0.255, 0.27, 0.285, and 0.3.  The color transformations include 
the effects of $\alpha$-enhancement but all assume Y$\sim$0.25.

\section{Age Constraints from the CMD}
The analysis performed in this section uses the color-magnitude
diagram (CMD) of 47 Tuc from the ACS Survey of Galactic Globular
Clusters (Sarajedini et al. 2007; Anderson et al. 2008).  The 
data have been culled (by removing stars with large photometric 
errors, see $\S$7 of Anderson et al. 2008) in order to more clearly 
delineate the main sequence but the final CMD still contains $\sim$50,000 stars.

In order to simplify the analysis, the apparent distance modulus 
and reddening were fixed to those values listed in Table \ref{tab1}.
Hence the only uncertainties in the ages derived from isochrone 
fitting are due to the inherent scatter in the data and the allowed 
range of metallicities.

\begin{figure}[htbp]
\vspace*{4.0 cm}
\includegraphics[width=6in]{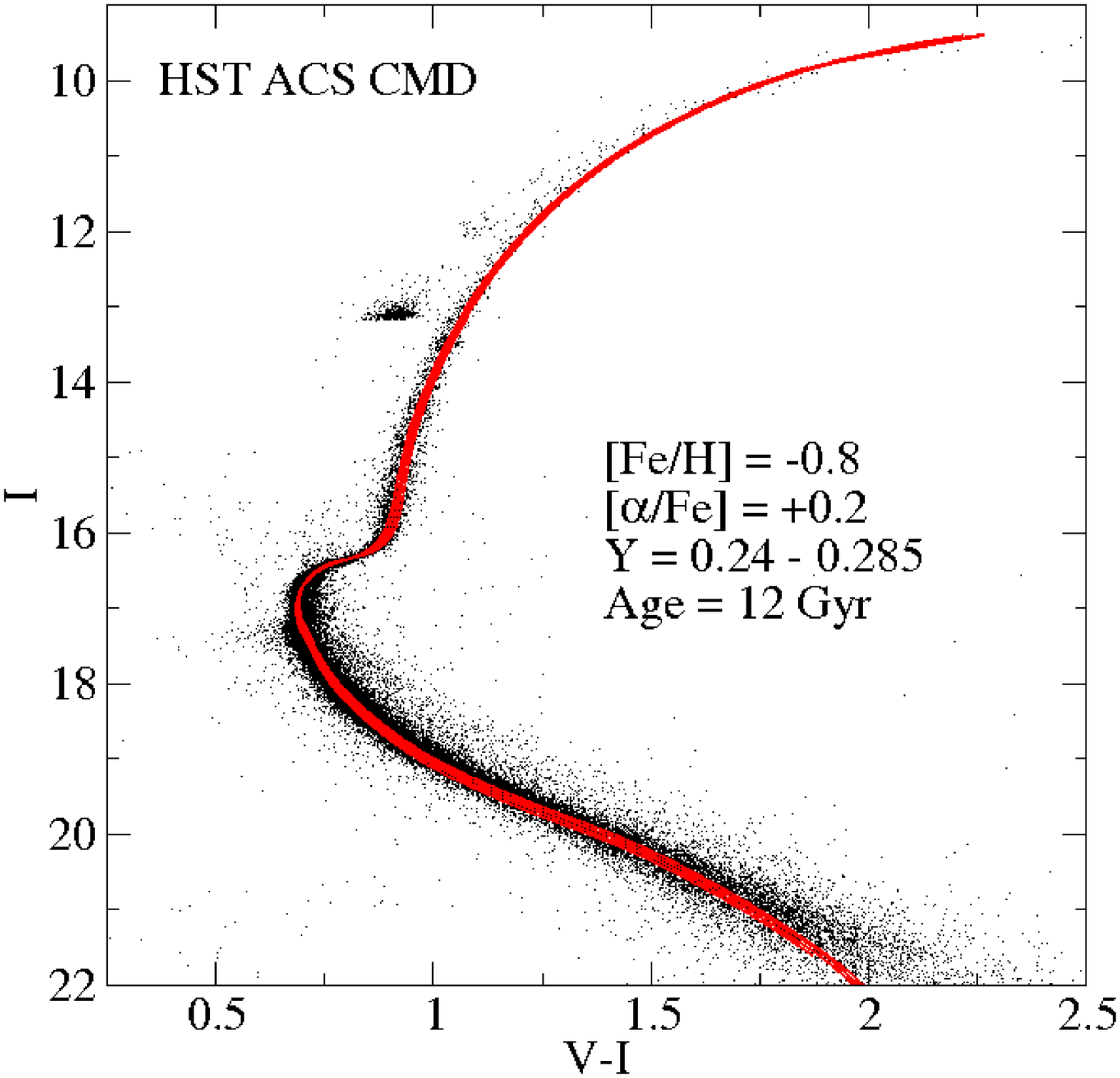} 
\caption{The effect of varying Y at fixed [Fe/H] and [$\alpha$/Fe].
Shown are isochrones with Y=0.24, 0.255, 0.27, and 0.285.}
\label{CMDY}
\end{figure}

\begin{figure}[htbp]
\vspace*{4.0 cm}
\includegraphics[width=6in]{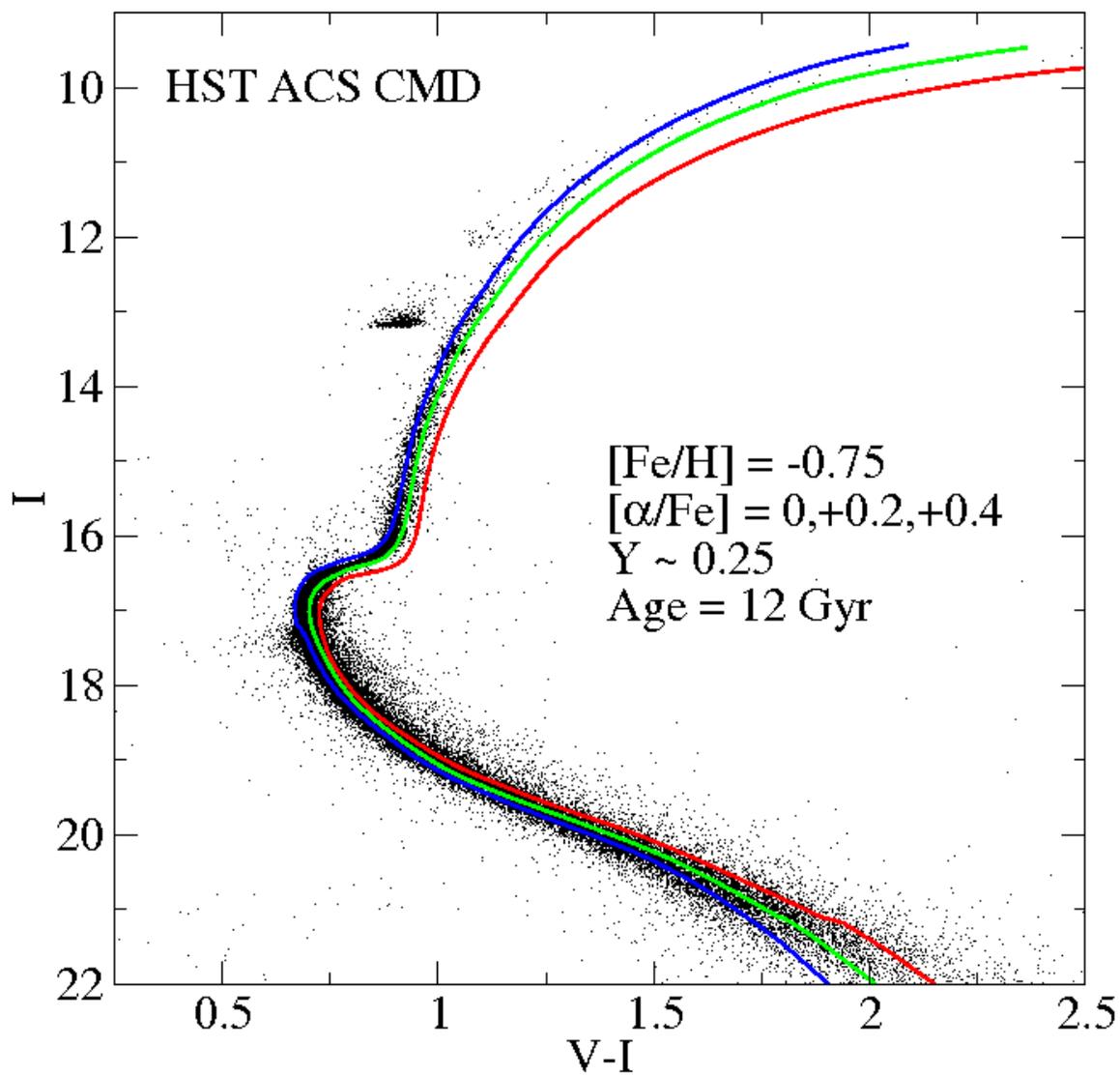} 
\caption{The effect of varying [$\alpha$/Fe] at fixed [Fe/H] and Y.}
\label{CMDA}
\end{figure}

\begin{figure}[htbp]
\vspace*{4.0 cm}
\includegraphics[width=6in]{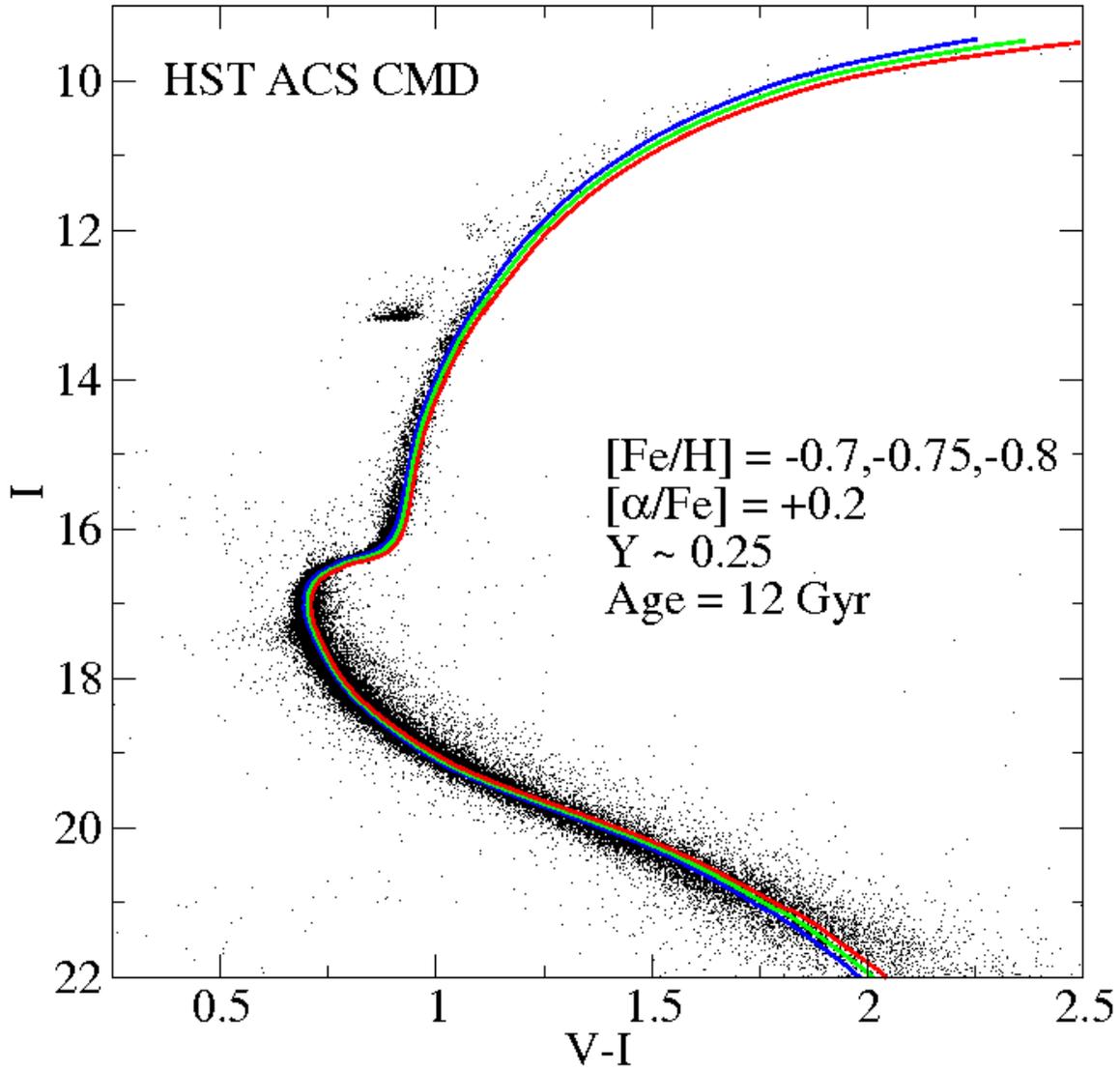} 
\caption{The effect of varying [Fe/H] at fixed [$\alpha$/Fe] and Y.}
\label{CMDZ}
\end{figure}

The basic results from fitting isochrones to the CMD are as follows:
(i) the main sequence and red giant branch impose little or no 
constraint on Y variations at the $\sim$0.05 level (see Figure \ref{CMDY}); (ii) the models 
favor a level of [$\alpha$/Fe] that is lower than the spectroscopic 
value (see Figure \ref{CMDA}) and indeed appear to rule 
out [$\alpha$/Fe]=+0.4; and (iii) the range of possible [Fe/H] values 
listed in Table 1 all give acceptable fits to the CMD assuming the 
preferred value of [$\alpha$/Fe] is used (see Figure \ref{CMDZ}).

Figure \ref{isoage} shows how age and [Fe/H] (assuming [$\alpha$/Fe]=+0.2 and Y$\sim$0.25) are
correlated.  The solid line is the best fit value as a function of 
[Fe/H] and the dashed lines represent the error bars that arise only
from the finite width of the subgiant branch.  A value of Y=0.255 was 
adopted for the plot but any Y value within the assumed range would give 
a similar result.  The isochrone fits to the CMD yield an age of 
11.5$\pm$0.75 Gyr where the uncertainty due to the fitting procedure is
$\sim$0.3 Gyr and the rest is due to $\sim$0.05 dex uncertainty in [Fe/H].

\section{Age Constraints from the Binary V69}
A careful analysis by Thompson et al. (2009) of spectroscopic 
radial velocity measurements and photometric light curves 
provides high precision estimates of the fundamental parameters of V69,
see Table \ref{tab2}. While the masses and radii are measured to 
better than 1\%, the luminosities are measured to no better than $\sim$10\% 
at present (though additional near-infrared light curves will substantially 
improve this number).

\begin{table}
\begin{center}
\caption{Parameters of V69 from Thompson et al. (2009)}
\label{tab2}
\begin{tabular}{|c|c|c|}\hline 
{\bf Parameter} & {\bf Primary} & {\bf Secondary} \\ \hline
$M/M_{\odot}$ & 0.8762$\pm$0.0048 & 0.8588$\pm$0.0060 \\ \hline
$R/R_{\odot}$ & 1.3148$\pm$0.0051 & 1.1616$\pm$0.0062 \\ \hline
$L/L_{\odot}$ & 1.94$\pm$0.21     & 1.53$\pm$0.17     \\ \hline
\end{tabular}
\end{center}
\end{table}

In order to measure the age of 47 Tuc using V69, individual stellar
evolution tracks were computed for six different masses: three for
each star, encompassing the central value and quoted uncertainties
in Table \ref{tab2}.  These models were calculated for the same composition grid
as the isochrones described in section 2.  The mass-radius (M-R) and 
mass-luminosity (M-L) relations in the tracks were used to determine 
the range of age for which each track lay within the error bars for 
the corresponding parameter (R or L) of each component.  We make two 
important general comments: (i) the ages derived from each relation 
are in excellent agreement between the two stars and (ii) the 
agreement between the two methods in a given star are quite sensitive 
to Y.  For either star, the maximum Y value for which the M-R and M-L 
results overlap increases with [Fe/H].

The results of the V69 age analysis are presented in the Figure \ref{bin}.
The solid lines indicate how the measured age varies with [Fe/H] for
the Y value listed on the figure.  The dashed lines above and below
the line for Y=0.255 indicate the combined uncertainties in M, R, and 
L, with $\sim$0.5 Gyr due to the luminosity uncertainty alone.  Figure 
\ref{bin} clearly shows that the age-[Fe/H] relation is complimentary 
to that of the CMD method (Figure \ref{isoage}).  Of equal importance is 
the fact that since the masses of the stars are known, there is a 
significant sensitivity to the He abundance that is lacking in the CMD 
method.

\section{Combined Constraints}
It was demonstrated in $\S$3 that the CMD age-[Fe/H] relation 
(assuming that both the distance and reddening are well-constrained) is 
linear for --0.8 $<$ [Fe/H] $<$ --0.7 and that age decreases as 
[Fe/H] increases.  Conversely, the result from $\S$4 showed
that while the age-[Fe/H] relation derived from V69 was also linear
over the range of [Fe/H] considered, the age increased as [Fe/H]
increased.  Combining the two methods is therefore quite powerful 
because the slopes have opposite signs.  The region 
where the two overlap has a finite size and gives preferred values
for age, [Fe/H], and Y simultaneously, though  ultimately the size of 
the overlap region is dominated by the uncertainties in both methods.

\begin{figure}[htbp]
\vspace*{4.0 cm}
\includegraphics[width=6in]{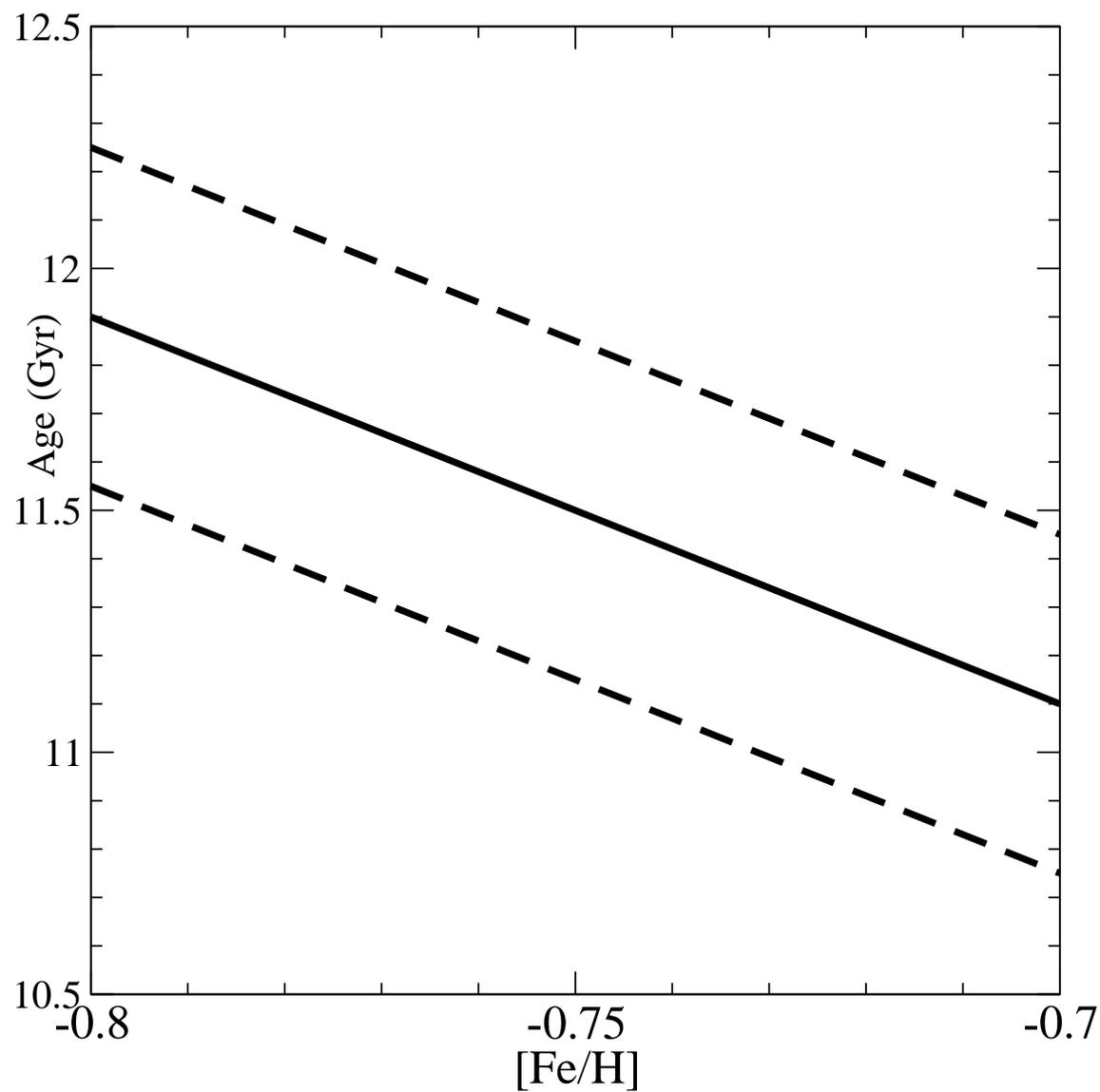} 
\caption{Isochrone fitting results in the age--[Fe/H] plane.  The
solid line represents the best fits while the dashed lines indicate
the uncertainties due only to the width of the subgiant branch.}
\label{isoage}
\end{figure}

\begin{figure}[htbp]
\vspace*{4.0 cm}
\includegraphics[width=6in]{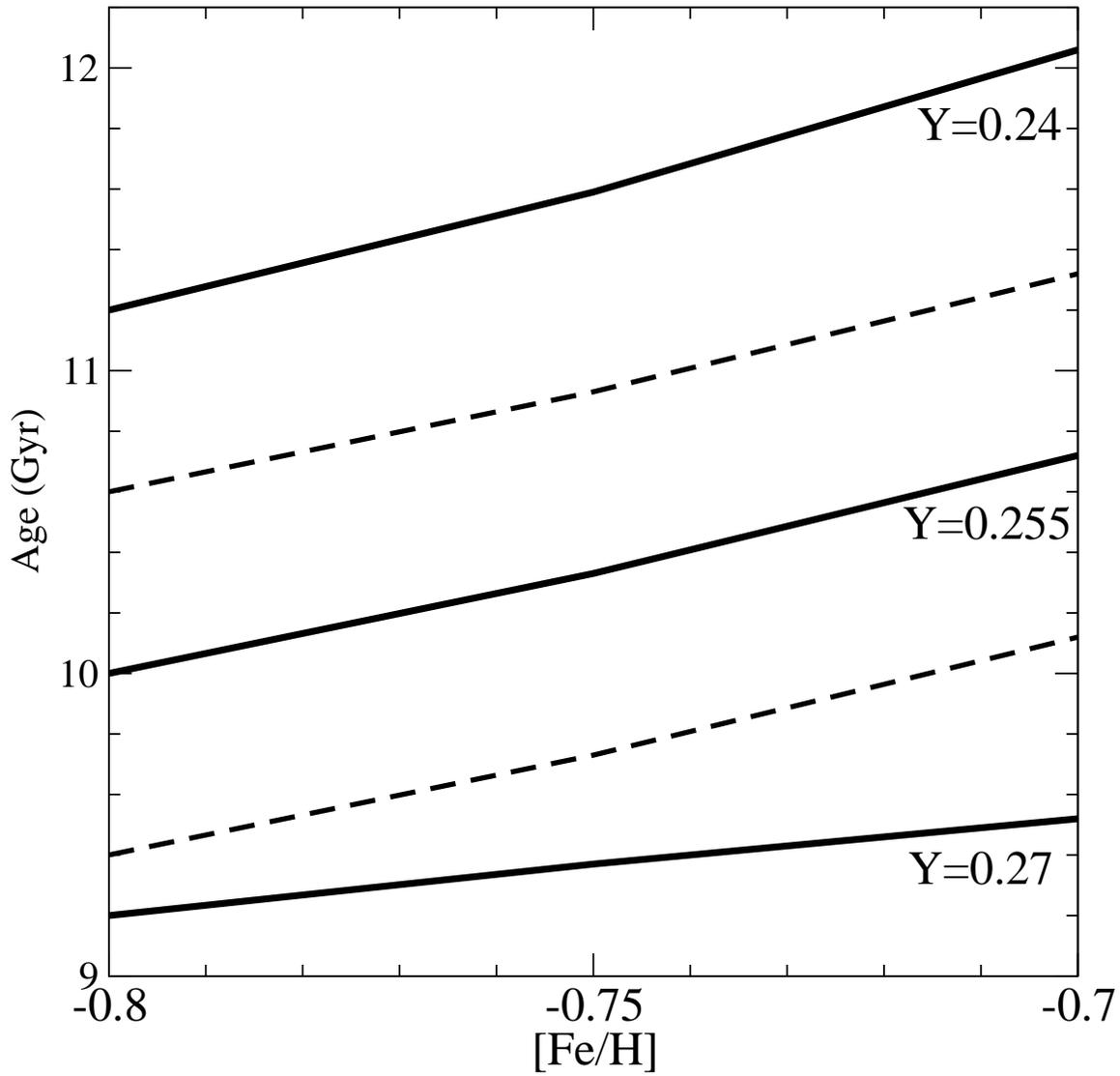} 
\caption{V69 results in the age--[Fe/H] plane.  The solid lines
correspond to results from the labeled Y values. The dashed lines
about the Y=0.255 line indicate the errors for that particular case
which are typical of all cases.}
\label{bin}
\end{figure}

\begin{figure}[htbp]
\vspace*{4.0 cm}
\includegraphics[width=6in]{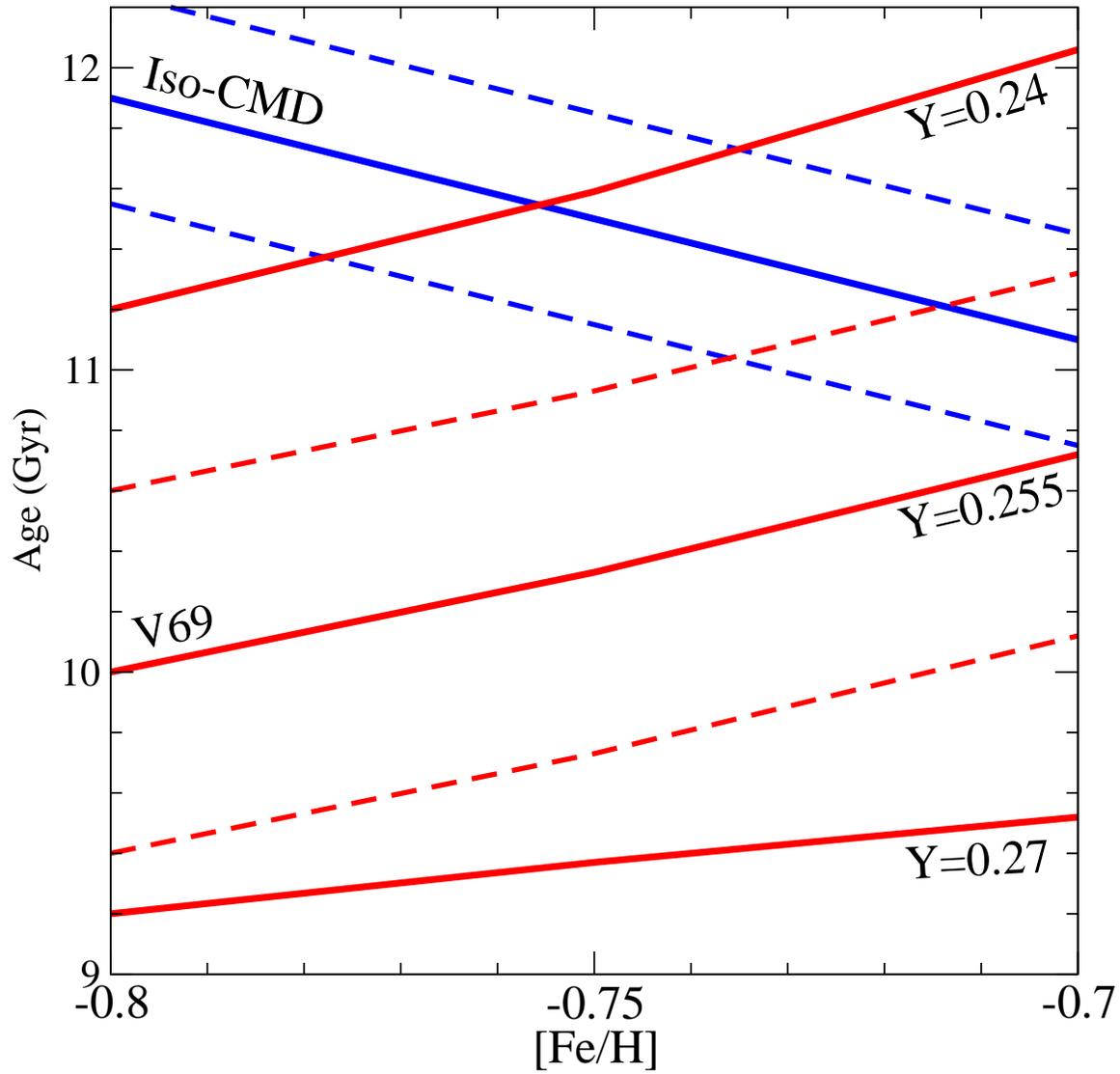} 
\caption{Combined isochrone-CMD and V69 results in the age--[Fe/H] plane.}
\label{comb}
\end{figure}

Figure \ref{comb} presents the results given in Figures \ref{isoage}
and \ref{bin} together so that the situation described in the preceding 
paragraph can be clearly seen.  The figure indicates that the binary and 
CMD methods agree best if the He abundance is low (Y $\sim$ 0.24) while a 
value of Y $\sim$ 0.25 would have been more preferable given the constraints
from WMAP and Big Bang nucleosynthesis, see Spergel et al. (2003).

If the bias towards [$\alpha$/Fe] = +0.2 derived from isochrone fits 
to the CMD in section 2 is relaxed and the spectroscopic value (+0.4)
is adopted instead, then the combined results favor Y $\sim$ 0.25.
However, it is difficult to reconcile the larger [$\alpha$/Fe] ratio
given the precise reddening.

Further improvements to the distance and luminosity of V69 from
near infrared light curves will significantly reduce the uncertainties
and provide a much more stringent test of the stellar evolution models'
ability to simultaneously fit the components of V69 and yield an age
consistent with the turnoff age.  As such, the ongoing CASE project 
represents the most exacting test of stellar evolution models in  
globular clusters to date and is certain to provide invaluable insights
into the ages--and other aspects--of such systems. 

\acknowledgments
Support for AD is provided by CITA and an NSERC grant to Don VandenBerg.
Research of JK is supported by the Foundation for Polish Science through
the grant MISTRZ. IBT is supported by NSF grant AST-0507325.


\begin{thebibliography}{}

\bibitem[Anderson et al. (2008)]{a08}
{Anderson, J. et al.} 2008, \textit{AJ}, 135, 2114

\bibitem[Chaboyer et al. (2001)]{ch01}
{Chaboyer, B. et al.} 2001, \textit{ApJ}, 562, 521

\bibitem[Dotter et al. (2007)]{d07}
{Dotter, A. et al.} 2007, \textit{AJ}, 134, 376

\bibitem[Dotter et al. (2008)]{d08}
{Dotter, A. et al.} 2008, \textit{ApJS}, 178, 89

\bibitem[Gratton et al. (2003)]{g03}
{Gratton, R. G. et al.} 2003, \textit{A\&A}, 408, 529

\bibitem[Koch \& McWilliam (2008)]{km08}
{Koch, A. \& McWilliam, A.} 2008, \textit{AJ}, 136, 518

\bibitem[Liu \& Chaboyer (2000)]{lc00}
{Liu, W. M. \& Chaboyer, B.} 2000, \textit{ApJ}, 544, 818

\bibitem[McWilliam \& Bernstein (2008)]{mb08}  
{McWilliam, A. \& Bernstein, R. A.} 2008, \textit{ApJ}, 684, 362

\bibitem[Percival et al. (2002)]{p02}
{Percival, S. M. et al.} 2002, \textit{ApJ}, 573, 174

\bibitem[Salaris et al. (2007)]{s07}
{Salaris, M. et al.} 2007, \textit{A\&A}, 476, 243

\bibitem[Salaris \& Weiss (2002)]{sw02}
{Salaris, M. \& Weiss, A.} 2002, \textit{A\&A}, 388, 492

\bibitem[Sarajedini et al. (2007)]{sa07}
{Sarajedini, A. et al.} 2007, \textit{AJ}, 133, 1658

\bibitem[Schlegel et al. (1998)]{sch98}
{Schlegel, D. J. et al.} 1998, \textit{ApJ}, 500, 525

\bibitem[Spergel et al. (2003)]{spe03}
{Spergel, D. N. et al.} 2003, \textit{ApJS}, 148, 175

\bibitem[Thompson et al. (2009)]{th09}
{Thompson, I. B. et al.} 2009, \textit{AJ}, submitted

\bibitem[VandenBerg (2000)]{vdb00}
{VandenBerg, D. A.} 2000, \textit{ApJS}, 129, 315

\end{thebibliography}
\end{document}